\documentclass[twocolumn,showpacs,preprintnumbers,amsmath,amssymb]{revtex4}
\usepackage[dvips]{color}
\usepackage{graphicx}
\usepackage{dcolumn}
\usepackage{bm}
\begin{document}
\preprint{APS/123-QED}
\title{Coherent acceleration of material wavepackets in
modulated optical fields}
\author{F. Saif$^{1,2}$}
\email{saif@fulbrightweb.org}
\author{P. Meystre$^2$}
\affiliation{$^1$Department of Electronics, Quaid-i-Azam University,
Islamabad 45320, Pakistan.\\
$^2$Department of Physics, The
University of Arizona, Tucson, AZ 85721, USA. }
\date{\today}
\begin{abstract}
We study the quantum dynamics of a material wavepacket bouncing
off a modulated atomic mirror in the presence of a gravitational
field. We find the occurrence of coherent accelerated dynamics for
atoms beyond the familiar regime of dynamical localization. The
acceleration takes place for certain initial phase space
data and within specific windows of modulation strengths. The
realization of the proposed acceleration scheme is within the
range of present day experimental possibilities.
\end{abstract}

\pacs{03.75.-b, 39.20.+q, 03.65.-w, 47.52.+j, 05.45.-a}

\maketitle

Enrico Fermi, in his seminal paper 'On the origin of cosmic rays',
conjectured that ``cosmic rays are originated and accelerated
primarily in the interstellar space of the galaxy by {\it
collisions} against moving magnetic fields''~\cite{kn:fermi}.
Following this work, the possibility of accelerating particles
bouncing off oscillating surfaces~\cite{saifpla} was extensively
studied in accelerator physics, leading to the development of two
major models: The Fermi-Ulam accelerator, which deals with the
bouncing of a particle off an oscillating surface in the presence
of another fixed surface parallel to it; and the
Fermi-Pustyl'nikov accelerator, where the particle bounces off an
oscillating surface in the presence of gravity. In the case of the
Fermi-Ulam accelerator~\cite{kn:lieb,kn:lilico} it was shown that
the energy of the particle remains bounded and the unlimited
acceleration proposed by Fermi is absent \cite{zaslavskii}. In the
Fermi-Pustyl'nikov accelerator, by contrast, there exists a set of
initial data within specific domains of phase space that result 
in trajectories speeding up to
infinity. 

In recent years, the efficient transfer of large momenta to
laser-cooled atoms has become an important problem for a number of
applications such as atom interferometry and the development of
matter-wave based inertial sensors in quantum metrology. Possible schemes of
matter-wave acceleration have been proposed and studied. For
example, a spatially periodic optical potential applied at
discrete times to an atom~\cite{modwave} was found to accelerate
it in the presence of a gravitational field~\cite{accelm}. This
$\delta$-kicked accelerator operates for certain sets of initial
data that originate in stable islands of phase space.

In this Letter, we propose an experimentally realisable new
technique to accelerate a material wavepacket in a coherent
fashion. It consists of an atom optics version of the
Fermi-Pustyl'nikov accelerator ~\cite{Saif 1998} where a cloud of
ultracold atoms falling in a gravitational field bounces off a
spatially modulated atomic mirror. This scheme is different from
previous accelerator schemes in the following ways: {\it i}) The
regions of phase space that support acceleration are located in
the mixed phase space rather than in the islands of stability (or
nonlinear resonances); {\it ii}) The acceleration of the
wavepacket is coherent; {\it iii}) It occurs only for certain
windows of oscillation strengths.
    \begin{figure*}
    \includegraphics[scale=0.85]{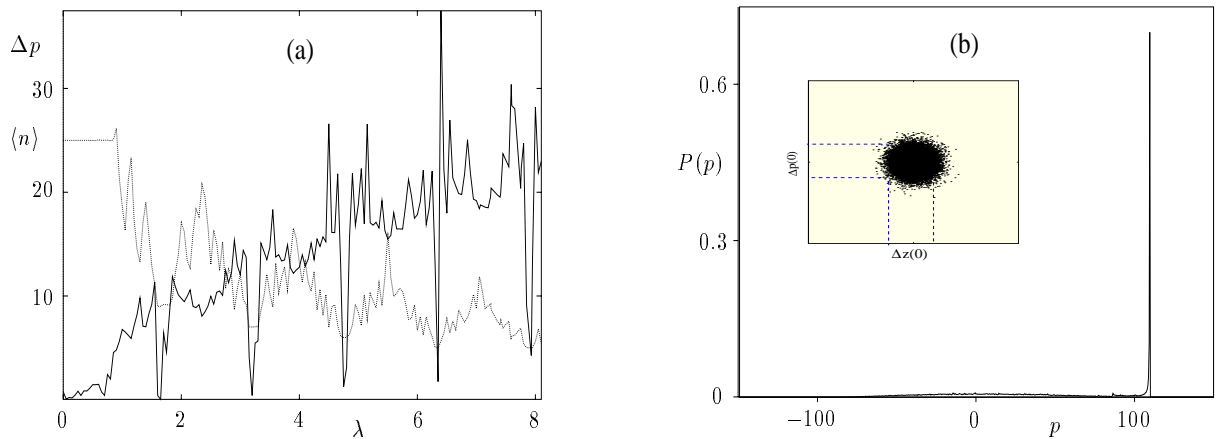}
    \caption{(a) Varience of the momentum distribution, $\Delta p$, as a
    function of the modulation strength $\lambda$, for an ensemble of
    particles initially in a narrowly peaked Gaussian distribution
    originating from the area of phase space that supports
    accelerated trajectories. The variance $\Delta p$
    (solid line) and average number of bounces $\langle n \rangle$
    (dotted gray line) are displayed after an evolution time $t = 300$,
    for an ensemble of 10000 particles. The initial distributions
    in momentum and in coordinate space are gaussians centered at $\bar z=0$
    and $\bar p=2\pi^2$ with $\Delta p(0) = \Delta z(0)=0.1$. The momentum
    distribution at $t=300$ is illustrated in (b) for $\lambda=3\pi/2$. The broad,
    barely visible background results from the tails of the initial
    gaussian distribution outside the area of phase space
    that supports accelerated trajectories.
    The numerical calculations correspond to Cesium atoms of mass $m=2.2
    \times 10^{-25}$Kg bouncing off an atomic mirror with an intensity
    modulation of $\epsilon=0.55$. The
    modulation frequencies extend to the megahertz range, and
    $\kappa^{-1}=0.55$ $\mu$m. } \label{one}
    \end{figure*}

Our starting point is the analysis by Saif {\it et al.}~\cite{Saif
1998} that establishes the dynamical localization of atoms in the
Fermi-Pustyl'nikov accelerator (or Fermi accelerator for short)
and shows a diffusive behavior both in the classical and the
quantum domains beyond the localization regime~\cite{kn:sten}. We
now extend these results to identify conditions leading to the
coherent acceleration of the atoms. We find clear signatures of
that behavior both in an ensemble of classical particles and for a
quantum wavepacket. In quantum mechanics, however, the Heisenberg
Uncertainty Principle restricts the phase-space size of the
initial atomic wavepacket which may result in the coherent
acceleration occurring on top of a diffusive background.

We consider a cloud of laser-cooled atoms that move along the
vertical $\tilde z$-direction under the influence of gravity and
bounce back off an atomic mirror~\cite{kn:amin}. This mirror is
formed by a laser beam incident on a glass prism and undergoing
total internal reflection, thereby creating an optical evanescent
wave of intensity $I(\tilde z)=I_0\exp(- 2k \tilde z)$ and
characteristic decay length $k^{-1}$ outside of the prism.

The laser intensity is modulated by an acousto-optic modulator as
\cite{kn:sten}
\begin{equation}
I(\tilde z,\tilde t) = I_0\exp(- 2k\tilde z
+\epsilon\sin\omega\tilde t),
\end{equation}
where $\omega$ is the frequency and $\epsilon$ the amplitude of
modulation. The laser frequency is tuned far from any atomic
transition, so that there is no significant upper-state atomic
population. The excited atomic level(s) can then be adiabatically
eliminated, and the atoms behave for all practical purposes as
scalar particles of mass $m$ whose center-of-mass motion is
governed by the one-dimensional Hamiltonian
\begin{equation}
\tilde{H}=\frac{\tilde{p}^2}{2m}+mg\tilde{z}
+\frac{\hbar\Omega_{\rm eff}}{4} e^{-2k\tilde{z}+
\epsilon\sin\omega\tilde t}, \label{ham}
\end{equation}
where $\tilde p$ is the atomic momentum along $\tilde{z}$ and $g$
is the acceleration of gravity.

We proceed by introducing the dimensionless position and momentum
coordinates $z\equiv\tilde{z}\omega^2/g$ and
$p\equiv\tilde{p}\omega/(mg)$, the scaled time
$t\equiv\omega\tilde t$, the dimensionless intensity
$V_0\equiv\hbar\omega^2\Omega_{\rm eff}/(4mg^2)$, the
``steepness'' $\kappa\equiv 2kg/\omega^2$, and the modulation
strength $\lambda\equiv\omega^2\epsilon/(2kg)$ of the evanescent
wave field, in terms of which the Hamiltonian takes the
dimensionless form
    \begin{equation}
    H=\left (\omega^2/mg^2 \right )\tilde{H}=\frac{p^2}{2}+z+\exp{[-\kappa(z-\lambda
    \sin t)]}.
    \end{equation}
When extended to an ensemble of non-interacting particles, the
classical dynamics obeys the condition of incompressibility of the
flow~\cite{kn:lieb}, and the phase space distribution
function $P(z,p,t)$ satisfies the Liouville equation
\begin{equation}
\left\{ \frac{\partial}{\partial { t}} + { p}
\frac{\partial}{\partial { z}} + {\dot p} \frac{\partial}{\partial
{ p}}\right\} P({z,p,t}) = 0,  \label{eq:reliou}
\end{equation}
where $\dot{{ p}}=-1 + \kappa V_0 \exp\left [-\kappa (z
-\lambda\sin t)\right]$ is the force on a classical particle.

In the absence of mirror modulation, the atomic dynamics is
integrable. For very weak modulations the incommensurate motion
almost follows the integrable evolution and remains rigorously
stable, as prescribed by the KAM theorem. As the modulation
increases, though, the classical system becomes chaotic.

In the quantum regime, the atomic evolution is determined by the
corresponding Schr\"odinger equation
\begin{equation}
ik^{\hspace{-2.1mm}-}\frac{\partial\psi}{\partial
t}=\left[\frac{p^2}{2} + z + V_0 \exp\left[-\kappa (z -\lambda\sin
t)\right]\right]\psi \label{dham}
\end{equation}
where $k^{\hspace{-2.1mm}-}\equiv\hbar\omega^3/(mg^2)$ is the
dimensionless Planck constant, introduced consistently with the
commutation relation $[z,p] ={\it i}(\omega^3 /mg^2) \hbar\equiv
{\it i}k^{\hspace{-2.1mm}-}$ for the dimensionless variables $z$
and $p$. We use Eqs. (\ref{eq:reliou}) and (\ref{dham}) to study
the classical and quantum mechanical evolution of an ensemble of
atoms in the Fermi accelerator.

For very short decay lengths $\kappa^{-1}$ and atoms initially far
from the mirror surface, we may approximate the optical potential
by an infinite potential barrier at the position $z=\lambda
\sin\omega t$. In that limit the atoms behave like falling
particles bouncing off a hard oscillating surface~\cite{kn:chen,
Saif 1998}.

Classical version of the problem~\cite{kn:pust} 
demonstrats the existence of a set of initial
conditions resulting in trajectories that accelerate without
bound. Specifically, the classical evolution of the Fermi
accelerator displays the onset of global diffusion above a
critical modulation strength $\lambda_l=0.24$ \cite{kn:lilico},
while the quantum evolution remains localized until a larger value
$\lambda_u$ of the modulation~\cite{Saif 1998, kn:benv,kn:oliv}.
Above that point both the classical and the quantum dynamics are
diffusive. However, for specific sets of initial conditions that
lie within phase space disks of radius $\rho$, accelerating
modes appear for values of the modulation strength $\lambda$
within the windows~\cite{kn:pust}
\begin{equation}
s\pi \leq \lambda < \sqrt{1+(s\pi)^2},  \label{eq:lcon}
\end{equation}
where $s$ can take integer and half-integer values for the
sinusoidal modulation of the reflecting surface considered here.

In order to compare the classical and quantum atomic dynamics
within these windows in the present situation we calculate the
width of the momentum distribution, $\Delta p\equiv
\sqrt{\langle{p^2}\rangle-{\langle p\rangle}^2}$, as a function of
the modulation strength $\lambda > \lambda_u$. In the classical
case, we consider an ensemble of particles with a gaussian initial
phase-space distribution $P(z,p,0)$ centered in a region of phase
space that supports unbounded acceleration, and record the
dispersion in momentum of that ensemble after a fixed propagation
time. The corresponding quantum problem is treated by directly
solving the Schr{\"o}dinger equation for an initial gaussian
wavepacket of zero mean momentum.

Classically, the momentum space variance $\Delta p$ remains small
and almost constant for very small modulation strengths, which
indicates the absence of diffusive dynamics, but starts to
increase linearly~\cite{Saif 1998} as a function of the modulation
strength for larger values of $\lambda$. As $\lambda$ is further
increased, though, we find that the diffusion of the ensemble is
sharply reduced for modulation strengths within the acceleration
windows of Eq.~(\ref{eq:lcon}), see Fig.~\ref{one} (a). Following
each window the momentum dispersion grows again approximately
linearly with $\lambda$. We interpret the sharply reduced value of
the dispersion as resulting from the non-dispersive, coherent
acceleration of the atomic sample above the atomic mirror which is
illustrated in Fig.~\ref{one} (b) and absent otherwise.

In the quantum case the Heisenberg Uncertainty Principle imposes a
limit on the smallest size of the initial wavepacket. Thus in
order to form an initial wavepacket that resides entirely within
regions of phase space leading to a coherent and dispersionless
acceleration, appropriate value of the effective Planck constant
is to be evaluated, for example, by controlling the frequency
$\omega$~\cite{tobe}. For broad wavepackets, the situation that we consider in
this Letter, the coherent acceleration manifests itself instead,
both in the classical and the quantum cases, as regular spikes in
the marginal probability distributions $P(p,t)= \int dx P(x,p,t)$
and $P(x,t)= \int dp P(x,p,t)$, which are absent otherwise.

\begin{figure}[t]
\begin{center}
\includegraphics[scale=0.4]{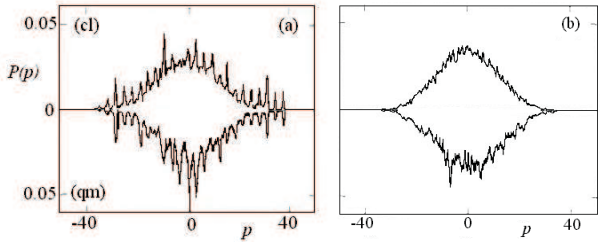}
\end{center}
\caption{Mirror images of the classical and quantum mechanical momentum distributions,
$P(p)$, plotted for (a) $\lambda=1.7$ and (b) $\lambda=2.4$, 
after a propagation time $t=500$. The
spikes in the momentum distribution for $\lambda=1.7$ are a
signature of coherent accelerated dynamics. The initial width of the momentum distribution is $\Delta p=0.5$. } \label{four}
\end{figure}

\begin{figure}
\begin{center}
\includegraphics[scale=.75]{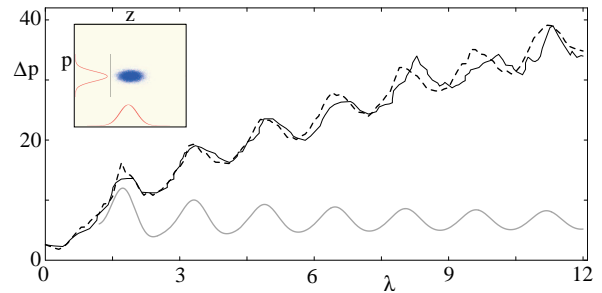}
\end{center}
\caption{Momentum variance $\Delta p$ as a function of $\lambda$ for an atomic de Broglie wave (thick line) and an ensemble of particles (dashed line) initially in a gaussian distribution after a scaled propagation time $t=500$. The initial probability distributions, shown in inset, have variances $\Delta p=0.5$ and $\Delta z=1$. The gray line shows scaled diffusion coefficient $D_{\lambda}/D_0$ as a
function of $\lambda$ (arbitrary units).} \label{two}
\end{figure}

\begin{figure}[t]
\begin{center}
\includegraphics[scale=0.55]{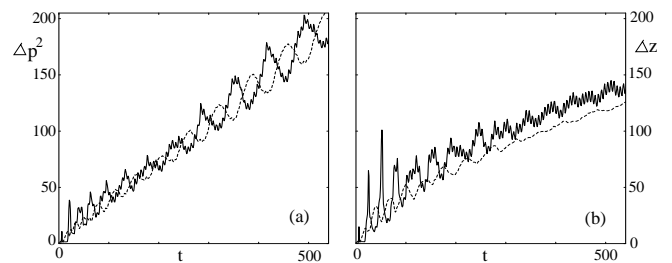}
\end{center}
\caption{(a) Square of the momentum variance (dark lines) and coordinate space 
variance (gray lines) as a function of time for $\lambda=1.7$. The coherent acceleration results in a breathing of the atomic wave packet, as evidenced by the out-of-phase oscillations of the variances.
(b) Dynamics for $\lambda=2.4$, a modulation strength that does not result in coherent acceleration. Note the absence of breathing in that case. The two values, $\lambda=1.7$ and $\lambda=2.4$, correspond to the first maxima and first minima in Fig.~\ref{two}. Same parameters as in Fig.~\ref{two}.} \label{three}
\end{figure}

Figure~\ref{four} illustrates the marginal probability distribution
$P(p,t)$ in momentum space for (a) $\lambda=1.7$ and (b) $\lambda=2.4$,
both in the classical and the quantum domains. In this example, the initial area of the
particle phase-space distribution is taken to be large compared to the
size of the phase-space regions leading to purely unbounded
dispersionless acceleration. The sharp spikes in time evolved momentum
distribution, $P(p,t)$ appear when the modulation strength satisfies the
condition of Eq.~(\ref{eq:lcon}), and gradually disappear as
it exits these windows. These spikes are therefore a
signature of the coherent accelerated dynamics. In contrast, the portions of the initial probability distribution originating from the regions of the phase space
that do not support accelerated dynamics undergo diffusive dynamics.

We can gain some additional understanding of the diffusive behavior from the
close mathematical analogy between the system at hand and the
kicked rotor model~\cite{Saif 1998}. It has been established
mathematically~\cite{Rechester} that for large modulation
strengths, the diffusive behavior of classical systems described
by the standard map displays a modulated growing behavior. For
large $\lambda$, the diffusion coefficient is given by
\begin{equation}
D_{\lambda}=D_0
\left(\frac{1}{2}-J_2(K)-J_1^2(K)+J_2^2(K)+J_3^2(K) \right),
\end{equation}
where $K=4\lambda$ and $D_0=K^2/2=8\lambda^2$~\cite{Rechester} and
$J_1$, $J_2$ and $J_3$ are first-, second- and third-order Bessel
functions. Recent experiments by Kanem {et al}~\cite{kanem} also
reports such a behavior for the $\delta$-kicked accelerator in
the case of large modulations.

A comparison between the classical behavior and the quantum
momentum dispersion as a function of $\lambda$ is illustrated in
Fig.~\ref{two}, while the scaled diffusion coefficient $D_\lambda/D_0$
is shown in arbitrary units. It is interesting to note that the dispersion exhibits maxima for oscillation strengths,
$\lambda_{m}= (s\pi + \sqrt{1+(s\pi)^2})/2$,
which reside at the center of the acceleration windows~\cite{saifpla},
indicating that those trajectories that do not correspond to the
phase space area supporting accelerated dynamics display maximum
dispersion instead.

From the numerical results of Fig.~\ref{four},
we conjecture that the spikes are well
described by a sequence of gaussian distributions separated by a
distance $\pi$, both in momentum space and coordinate space. 
We can therefore express the complete time-evolved
wavepacket composed of a series of sharply-peaked gaussian
distributions superposed to a broad background due to diffusive
dynamics, such that
\begin{eqnarray}
P(p)&=&{\cal N} e^{-p^2/4\Delta p^2}\sum_{n=-\infty}^{\infty}
e^{-(p-n\pi)^2/4\epsilon^2},
\end{eqnarray}
where $\epsilon << \Delta p$,
and $\cal N$ is a normalization constant.

Further insight in the quantum acceleration of the atomic wave packet is obtained by studying its temporal evolution. We find that within the window of acceleration the atomic wave packet displays a linear growth in the square of the momentum variance and in coordinate space variance as a function of time. 
Figure~\ref{three} illustrates that, for modulation strengths within the acceleration window, the growth in square of the momentum variance 
displays oscillations of increasing periodicity whereas the variance in
coordinate space follows with a phase difference of $180^o$. 
The out-of-phase oscillatory evolutions of $\Delta p^2$ and $\Delta z$ indicate a breathing of the
wavepacket and are a signature of the coherence in accelerated dynamics as it disappears outside. As a final point we note that outside of the acceleration window the linear growth in the square of the momentum variance, a consequence of normal diffusion, translates into a $t^{\alpha}$ law, with $\alpha < 1$ which is a consequence of anomalous diffusion. 

This work is supported in part by the US Office of Naval Research,
the National Science Foundation, the US Army Research
Office, the Joint Services Optics Program, the National
Aeronautics and Space Administration, and J. William Fulbright foundation.

\end{document}